\documentstyle[12pt,aasms4] {article}

\begin{document}

\title{On the Mass-Donating Star in GRS 1915+105}

\author{Stephen S. Eikenberry\altaffilmark{1} \& Reba M. 
Bandyopadhyay\altaffilmark{2,3}}

\altaffiltext{1}{Astronomy Department, Cornell University, Ithaca, NY 14853}
\altaffiltext{2}{E. O. Hulburt Center for Space Research, Naval Research 
Laboratory, Washington, DC 20375}
\altaffiltext{3}{NRL/NRC Research Associate}

\begin{abstract}

	We present an analysis of constraints on the companion mass
for the Galactic microquasar GRS 1915+105.  Using the known
inclination angle and stability of the jet axis, we can rule out
massive ($M_s \ga 3 M_{\sun}$) companions with orbital axes not
aligned with the jet axis.  For an aligned orbital axis, we constrain
the ratio of the stellar radius to the binary semi-major axis to be
$R_s/a <0.342$, based on the lack of X-ray eclipses.  We then show that
these constraints together with the X-ray luminosity, approximate
black hole mass, and observations of the X-ray absorbing column
density towards GRS 1915+105 rule out accretion via stellar wind from
a massive companion, implying that accretion occurs via Roche-lobe
overflow.  In that case, the constraint on $R_s/a$ rules out any
companion with $M_s > 19.4 M_{\sun}$ for a black hole mass of $M_{BH}
\simeq 30 M_{\sun}$.  The lack of significant X-ray reprocessing in
the infrared from GRS 1915+105 also provides constraints on the mass
and temperature range of a Roche-lobe-filling mass donor.

\end{abstract}


\section{Introduction}

	As the first Galactic source of superluminal jets, GRS
1915+105 has undergone intensive study during recent years.  Despite
this level of investigation, one of the system's key parameters -- the
approximate mass of the mass-donating star in the binary -- remains a
subject of controversy.  No binary modulation in the system, such as
eclipses, have been observed at any wavelength, preventing insights
based on orbital properties and evolution.  Mirabel et al., (1997)
observed near-infrared emission lines of \ion{He}{1} and Br$\gamma$,
which they interpreted as evidence for a massive Oe- or Be-type
companion star.  However, Castro-Tirado et al. (1996) and Eikenberry
et al. (1998) reported the observation of a weak, time-variable
\ion{He}{2} emission line (later confirmed by Marti et al., 2000).
Since normal Oe and Be stars do not typically doubly-ionize He, they
suggested that the emission lines arise in the accretion disk
surrounding the compact object, leaving the companion star type
unconstrained.  Furthermore, Eikenberry et al. (1998) observed the
emission lines changing in strength by factors of several on
timescales of $\sim 5-10$ minutes during jet-producing activity.  They
interpreted the constancy of the line profiles during these rapid
changes as further evidence of the accretion disk origin of the lines.

	More recently, Marti et al. (2000) have presented VLT spectra
of GRS 1915+105, in which they confirm the weak \ion{He}{2} emission
line and identify another emission line due to NaI ($2.21 \mu$m).
They interpret this feature as evidence that the mass-donating star in
GRS 1915+105 is a super-massive star, such as a luminous blue variable
(LBV) star.  However, they note that this interpretation does not seem
able to explain the line variability observed by Eikenberry et
al. (1998).

	In this paper, we will also address the nature of the
mass-donating star.  Based on the known inclination angle of the jet
axis, its stability over timescales of years, and the lack of X-ray
eclipses in GRS 1915+105, we place constraints on the ratio of the
mass-donating star's radius $R_s$ to the semi-major axis of the binary
system $a$.  We then show, based on estimates of the black hole mass
in GRS 1915+105, its X-ray luminosity, and the observed X-ray
absorption column towards GRS 1915+105, that the accretion in the
system cannot be provided by the wind of a massive star.  We then show
that for a $30 M_{\sun}$ black hole undergoing accretion via Roche
lobe overflow, the mass-donating star is constrained to have $M_s <
19.4 M_{\sun}$.  Next, we use limits on X-ray reprocessing on the
stellar surface to place further constraints on the mass-donating
star.  Finally, we discuss the implications of these results for GRS
1915+105 and present our conclusions.

\section{Geometry of GRS 1915+105}

	In the paper announcing the discovery of superluminal motion
from GRS 1915+105, Mirabel \& Rodriguez (1994) were able to constrain
the inclination angle of the jet axis to be $\theta = 70 \arcdeg \pm 2
\arcdeg$.  Furthermore, in the observed jet outbursts since then, this
angle has remained stable to within a few degrees on timescales of
years (e.g. Fender et al., 1999).  Thus, the orbital motion of the
compact object does not significantly alter the direction of the jet
axis.  This constancy can be provided by one of two possibilites:
either the mass ratio of the system is small ($q = M_s / M_{BH} \ll
1$), or else the orbital axis is aligned with the jet axis.  The first
case in and of itself strongly constrains the nature of the companion star.  
In the second case, further analysis is required to see if
the nature of the companion star can be constrained.

	The first component in this analysis is the lack of X-ray
eclipses in GRS 1915+105.  This object has been monitored by the Rossi
X-ray Timing Explorer (RXTE) All-Sky Monitor since early 1996, and
with a huge number of pointed observations using the RXTE Proportional
Counter Array (e.g. Muno et al., 2000).  It has also been extensively
studied using ASCA (e.g. Kotani et al., 2000), BeppoSAX (e.g. Feroci
et al., 2000), Granat-Sigma (Castro-Tirado, et al., 1994), and the
Compton Gamma-Ray Observatory (e.g. Harmon, et al., 1997).  Despite
this wealth of coverage, no X-ray eclipse has ever been observed from
GRS 1915+105.  For an orbit aligned with the jet axis at $\theta = 70$
degrees, this constrains the ratio of the star radius to the orbital
separation to be

$${R_s \over{a}} \leq \cos (70 \arcdeg) = 0.342$$

\section{Accretion via Stellar Wind}

\subsection{Limits on $\dot M$ and the Stellar Wind}

	If the mass-donating star in GRS 1915+105 is in fact an
early-type star, one possibility for providing the accreted mass is
through a stellar wind.  The accretion rate must be able to account
for the high X-ray luminosity observed from GRS 1915+105, with $L_x
\simeq 3 \times 10^{39} \ {\rm ergs \ s^{-1}}$ (e.g. Belloni et al.,
1997).  If we assume a canonical efficiency of 15\% for converting
accreted mass into radiated energy (e.g. Frank et al., 1992), we
arrive at an accretion rate of $\dot M \simeq 3 \times 10^{-7} \
M_{\sun} \ {\rm yr^{-1}}$.

	Wind accretion is known to be rather inefficent compared to
Roche lobe overflow -- only the wind material directed towards the
compact object is accreted.  More quantitatively, the accretion rate
from the wind $\dot M_{wind}$ is related to the stellar mass loss rate
$\dot M_s$ according to

$$\dot M_{wind} \simeq \onequarter \ \dot M_s \ q^{-2} \ (R_s /a)^2$$

The mass of the black hole in GRS 1915+105 is not well-determined.
However, Zhang, et al. (1997) and Nowak, et al. (1997) suggest $M_{BH}
\sim 30 M_{\sun}$ based on the 67 Hz quasi-periodic oscillation
observed from the system (Morgan et al., 1997).  Furthermore, by
requiring the peak X-ray luminosity to be less than the Eddington
luminosity, we find $M_{BH} > 23 M_{\sun}$ -- consistent with the
result of Zhang, et al. (1997).  Therefore, we adopt $M_{BH}= 30
M_{\sun}$ here.  For a super-massive star as suggested by Marti et
al., we have $M_s \ga 50 M_{\sun}$ (Humphreys \& Davidson, 1994),
giving $q \ga 1.6$.  For an accretion rate $\dot M \geq 3 \times
10^{-7} M_{\sun} \ {\rm yr^{-1}}$, we then arrive at $\dot M_s \ga 2.7
\times 10^{-5} M_{\sun} \ {\rm yr^{-1}}$.  This mass-loss rate is very
high -- even the LBV stars have an upper end to their mass loss
between outbursts of $\sim 10^{-5} M_{\sun} \ {\rm yr^{-1}}$
(e.g. Davidson \& Humphreys, 1997).  Since GRS 1915+105 has shown
repeated patterns of X-ray emission over timescales of several years,
implying steady accretion over this interval as compared to the
variable aperiodic outflows typical of LBV stars in outburst
(e.g. Humphreys \& Davidson, 1994), it seems likely that any LBV-type
mass-donating star is not in outburst.  Thus, the wind mass-loss rate
required to produce the observed X-ray emission is marginally
consistent with the mass-loss rates known to occur in such stars.

	The required wind mass-loss rate decreases with decreasing $q$
until $q \simeq {R_s \over{a}} = 0.34$, at which point essentially all
of the stellar wind not initially flowing away from the compact object
is captured.  For $q=0.34$, we have $\dot M_s \ga 1.2 \times 10^{-6}
M_{\sun} \ {\rm yr^{-1}}$ for a $\sim 10 M_{\sun}$ star.  This
mass-loss rate is again consistent with the rates inferred for the O/B
companions in wind-fed high-mass X-ray binaries (e.g. Frank et al.,
1992), though this stellar mass range is at the lower limit for such
stars.

\subsection{Limits from X-ray Absorption}

	The tremendous stellar wind required to provide the X-ray
luminosity of GRS 1915+105 should also have an impact on the observed
absorbing column density towards the X-ray source.  Given a minimum
stellar mass-loss rate from above and the wind velocity $v_{wind}$, we
can place a lower limit on the wind density as a function of distance
from the star's center $r$ according to

$$ \rho (r) = {\dot M_s \over{4 \pi r^2 \ v_{wind}}}$$

We take $v_{wind} = 10^3 \ {\rm km \ s^{-1}}$ here -- somewhat larger than
typical for these systems, making the above density a firm lower
limit.

	For this minimum $\dot M_s$, the ratio $R_s/a$ is fixed at
0.342.  Furthermore, for supermassive LBV stars, an upper limit on
their size is $R_s \la 1$ AU, giving an upper limit on the orbital
separation of $a_{max} = 2.9$ AU.  For a less-massive O/B-type star,
we have $R_s \la 0.1$ AU, giving $a_{max} = 0.29$ AU.  Given these
system parameters and assuming a uniform spherical wind, we can
calculate the line-of-sight column density through the wind for an
observer at a $20 \arcdeg$ angle above the orbital plane.  It is
important to note that the assumption of uniform spherical outflow is
not generally critical for the results of this calculation.  Rather,
we only require that the line of sight sample the typical density
regimes of any more-complicated outflow -- a requirement that will be
satisfied for all but the most contrived geometries in this case.
Based on this, we estimate the lower limit to the column density to be
$n_H \ga 1.5 \times 10^{23} \ {\rm cm^{-2}}$ for an LBV-type star and
$n_H \ga 7 \times 10^{22} \ {\rm cm^{-2}}$ for an O/B-type star when
the mass-donating star lies between the observer and the X-ray source.

	For comparison, the measured total column density towards GRS
1915+105 is $n_H \simeq 6 \times 10^{22} \ {\rm cm^{-2}}$
(e.g. Belloni et al., 1997)\footnote{This measurement assumes solar
abundances.  For enriched winds (commonly seen in these stars), the
apparent absorption would be even higher, increasing the contrast
between observations and the wind-accretion model.}, and the vast
majority of this absorption is attributable to the intervening ISM
(Chaty, et al., 1996).  Furthermore, as the orbital orientation
changes, any wind-induced absorption column would vary systematically,
dropping by a factor of $>5$ at opposition -- see Figure 1.  However,
spectra collected over long time intervals show no such variability in
absorption for GRS 1915+105(e.g. Muno et al., 1999).  So, even the
lowest estimates for stellar winds capable of providing the accreting
matter produce X-ray absorption greatly in excess of that observed.
Therefore, we conclude that a stellar wind cannot provide the
accretion observed in GRS 1915+105.

\section{Accretion via Roche-lobe Overflow}

\subsection{Maximum stellar mass}

	Roche-lobe overflow is generally a much more efficient mode of
mass-transfer than stellar winds, with essentially all of the
mass-loss transferring onto the compact object.  Thus, we have $\dot
M_s \simeq 3 \times 10^{-7} \ M_{\sun} \ {\rm yr^{-1}}$.  Moreover, as
Eggleton (1983) has shown, Roche lobe overflow requires that the mass
ratio $q$ and the ratio of the stellar radius to the binary separation
follow

$$ {R_s \over{a}} = {0.49 \ q^{2/3} \over{0.6 q^{2/3} + {\rm ln} (1+q^{1/3})}}$$

	For our upper limit of $R_s / a \la 0.342$ above, we arrive at
an upper limit of $q \la 0.647$, corresponding to a limit on the
stellar mass of $M_s \la 19.4 M_{\sun}$ (assuming $M_{BH} = 30
M_{\sun}$).  This upper limit is more than a factor of 2 below the
{\it lower} limit for supermassive stars of the type suggested by
Marti et al. (2000) as the companion for GRS 1915+105.

\subsection{Limits from X-ray reprocessing}

	We can also place limits on Roche-lobe-filling companion stars
to GRS 1915+105 due to the apparent lack of significant infrared (IR)
flux from reprocessed X-rays in the system (Eikenberry et al., 2000).
They place limits on the IR flux density due to reprocessing of
$\Delta F_{\nu} < 3 \times 10^{-26} \ {\rm ergs \ s^{-1} \ cm^{-2} \
Hz^{-1}}$ at $\nu = 1.4 \times 10^{14}$ Hz ($2.2 \mu$m), in the
presence of $\Delta L_x \simeq 10^{39} \ {\rm ergs \ s^{-1}}$.  For
thermal reprocessing on a blackbody of temperature $T$, we have a
reprocessing efficiency $\epsilon$ (ratio of the bolometric
reprocessed luminosity to the emitted X-ray luminosity) given by

$$\epsilon = { \ c^2 \ \sigma T^3 \over{2 \pi \nu^2 \ k}} {\Delta L_{\nu} 
\over{\Delta L_x}}$$

assuming the change in temperature is small compared to the
temperature ($\Delta T \ll T$), changes in the surface area of the
blackbody are negligible, and ${\Delta L_{\nu} = 4 \pi d^2 \Delta
F_{\nu}}$ where we take $d = 12.5$ kpc (Mirabel \& Rodriguez, 1994).
From simple geometry, we can see that $\epsilon$ should also have the
form

$$ \epsilon = (1- \eta) {\pi \ R_s^2 \over{4 \pi \ a^2}}$$

where $\eta$ is the X-ray albedo of the stellar surface (assumed to be
$\eta \simeq 0.5$ here -- e.g. Anderson, 1981).

	Given this relationship between $\epsilon$ and $R_s \over{a}$
at a given temperature $T$ and the relationship between $q$ and $R_s
\over{a}$ for Roche-lobe overflow, we can then calculate an upper limit
on $q$ as a function of stellar surface temperature.  Figure 2 shows
a plot of the parameter space (mass, temperature) allowed/excluded for
Roche-lobe-filling stars\footnote{Note that for stars with $M_s \la 3
M_{\sun}$ ($q \la 0.1$), the constraint on the alignment of the
orbital axis with the jet axis may be relaxed, so that the stellar
temperature is not constrained by this method.} (assuming $M_{BH} = 30
M_{\sun}$).

	One important consideration here is that we have ignored the
possible effects of the accretion disk casting an X-ray shadow on the
star, thus reducing the amount of reprocessing on the stellar surface.
However, this intercepted radiation does not simply disappear --
rather it is reprocessed on the disk itself.  While the efficiency of
this reprocessing depends on the temperature and albedo of the disk
region intercepting the radiation, the star subtends only an azimuthal
range of $2 R_s \over{a}$ radians as seen from the center of the
accretion disk, while the disk extends a full $2 \pi$ radians.  Thus,
if the disk intercepts any fraction of the X-ray light that would
otherwise reach the star, it will in fact intercept at least $\pi a
\over{R_s}$ times that amount of light over its total azimuthal range.
Given our limit of ${R_s \over{a}} <0.342 $, this means that as long
as the disk reprocessing efficiency is at least 10\% of the stellar
reprocessing efficiency per unit area, any disk shadowing would
actually {\it increase} the total amount of reprocessed light over the
above estimate, and thus tighten the constraint given the upper limits
on reprocessed radiation.  Therefore, we consider the constraints in
Figure 2 to be fairly robust against the effects of disk shadowing.

\section{Discussion}

\subsection{The companion star mass}

	As shown above, any reasonable parameters for accretion via
stellar wind produce large X-ray absorption which is inconsistent with
observations of GRS 1915+105.  Furthermore, for accretion via
Roche-lobe overflow, we can rule out any star with $M_s > 25 M_{\sun}$
as well as all but the hottest stars with $M_s \ga 3 M_{\sun}$.  Thus,
a super-massive companion, as suggested by Marti et al. (2000), does
not seem possible for GRS 1915+105.

	The likeliest ``normal'' candidates for the mass-donating star
are either a main-sequence or giant (luminosity class III) early-B
star, or a low-mass ($M_s \la 3 M_{\sun}$) star.  For the expected
stellar radii and masses of massive stars and assuming accretion via
Roche lobe overflow, we would then expect binary periods for massive
donor stars in GRS 1915+105 to range over $\sim 10 - 15$ days.  For
low-mass companions, the range of expected binary periods is very
poorly constrained.

	Given the $\dot M \geq 3 \times 10^{-7} M_{\sun} \ {\rm
yr^{-1}}$ accretion rate required to produce the X-ray luminosity, one
might expect a low-mass star to imply a short lifetime for the
``active'' phase of GRS 1915+105 -- $\tau \sim M/{\dot M} \simeq
10^{6}-10^7$ years.  However, while GRS 1915+105 has been active for most
of the last 8 years, we know it was quiescent for several decades
prior to this.  Therefore, the duty cycle for active mass transfer in
this system may be $\ll 1$, so that the lifetime can be considerably
larger than the above estimate.

\subsection{Whence the IR emission lines?}

	The primary motivation for the initial suggestion of a
super-massive companion by Marti et al. (2000) was the IR emission
line spectrum.  The stars allowed under the above analysis could not
produce sufficient ionizing radiation to explain the observed IR
spectrum from GRS 1915+105.  However, it is important to note that the
IR lines observed from LBV and other such stars are {\it not} produced
directly in the stellar photosphere/chromosphere.  Rather, they are
the result of the dense stellar wind being irradiated by UV emission
from the luminous hot photosphere.

	In the case of GRS 1915+105, we {\it know} that the system
contains a hot, highly-luminous blackbody -- the inner accretion disk,
which we observe directly in X-rays, should also produce tremendous
ionizing UV luminosity.  Furthermore, the outer portions of such a
disk are widely considered to be the sources of strong winds, and we
are most interested in this particular system due to its strong
outflows.  Recent X-ray spectral analysis by Kotani et al. (2000)also
indicate the presence of such a disk wind.  Therefore, we hypothesize
that the emission spectrum observed by Marti et al. (2000) (as well as
others previous to them) derives not from a super-massive companion
star, but from the accretion disk itself.

	This latter explanation also fits well with the observations
of Eikenberry et al. (1998).  The observed rapid variability of the IR
emission lines is correlated with flaring behavior in GRS 1915+105 in
a manner suggesting radiative pumping.  Since the line profiles are
constant, the line-emitting region must be uniformly irradiated by the
flares which appear to come from the inner accretion disk.  This is
not possible for a line emission from the envelope of a massive
companion -- the star itself will shield a considerable fraction of
its envelope from the flare.  However, if the lines arise from the
disk, the line-emitting region is axisymmetric about the flaring
region, guaranteeing uniform irradiation and maintaining the constancy
of the emission line profiles.

\subsection{The black hole mass}

	A key assumption for much of the above analysis is the mass of
the black hole in GRS 1915+105.  As noted above, both arguments based
on the Eddington luminosity and (more speculative) interpretations of
quasi-periodic oscillations imply $M_{BH} \simeq 30 M_{\sun}$.  This
estimate would make the black hole in GRS 1915+105 the most massive of
the ``stellar-mass'' black hole candidates in the Galaxy -- roughly 3
times more massive than the black hole in Cyg X-1.  Based on the above
arguments, $M_{BH} = 30 M_{\sun}$ eliminates the possibility of a
super-massive companion star ($M_s \ga 50 M_{\sun}$) as suggested by
Marti et al. (2000).  Furthermore, in order to allow a companion
consistent with even the lowest portion of the LBV mass range would
require $M_{BH} \ga 75 M_{\sun}$.

\section{Conclusions}

	We have presented an analysis of the possible companion star
mass ranges for GRS 1915+105, and come to the following conclusions:

\begin{itemize}

\item Based on the $70 \arcdeg$ inclination angle and stability of the
jet axis, we find that either the mass ratio of the system is $q \ll
1$, or the orbital axis is aligned with the jet axis.

\item Based on the lack of X-ray eclipses, for massive companions ($q
\ga 0.1$), the radius of the star must be $R_s < 0.342 a$, where $a$ is
the semi-major axis of the orbit.

\item Given these constraints and the X-ray luminosity of GRS
1915+105, all reasonable parameters assuming accretion via a stellar
wind produce an X-ray aborption in the wind which is incompatible with
observations.  This indicates that accretion must occur via Roche-lobe
overflow in GRS 1915+105.

\item For Roche-lobe overflow, the constraint on $R_s/a$ requires the
companion star to have $M_s < 19.4 M_{\sun}$ (assuming $M_{BH} = 30
M_{\sun}$).

\item Based on limits for IR flux from reprocessed X-ray in GRS
1915+105 (Eikenberry et al., 2000), any companion star with $3
M_{\sun} < M_s < 25 M_{\sun}$ must have a surface temperature $T \ga
1-2 \times 10^4$ K.  The ``normal'' main sequence and giant stars
which fall into this range are early-B stars.

\item The IR emission lines do not arise in the envelope of a massive
companion, but instead are likely to be produced by the outer
accretion disk and/or a disk wind being irradiated by the hot inner
accretion disk.

\end{itemize}

\acknowledgements SE is supported in part by an NSF Faculty Early
Career Development (CAREER) award (NSF-9983830) and NASA grant
NAG5-9178.  This work was performed while RMB held a National Research
Council Research Associateship Award at NRL.

\vfill \eject

\begin{figure}
\vspace*{140mm}
\includegraphics{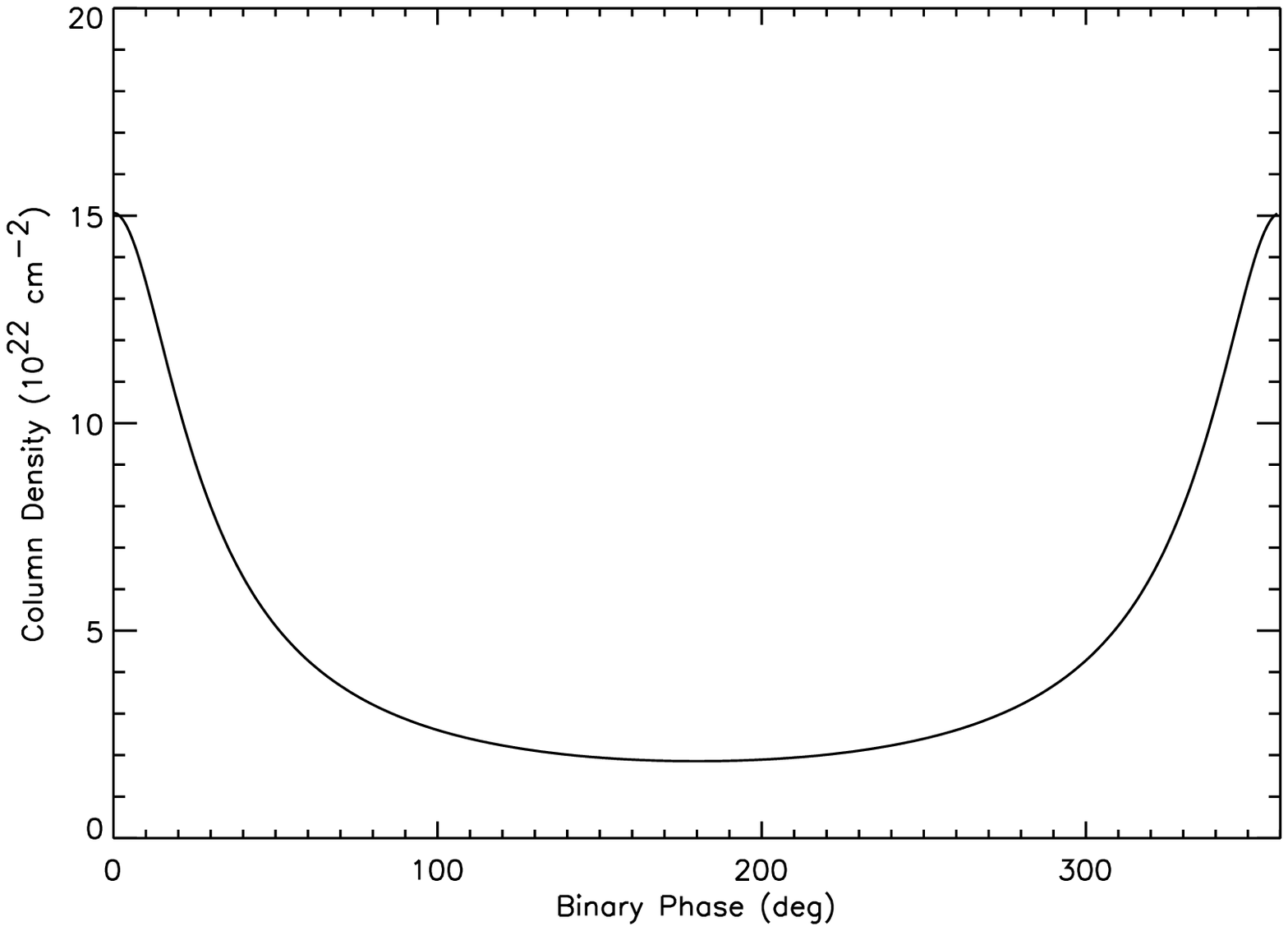}
\caption{Orbital variation of observed X-ray absorption column density
for wind-fed accretion from an LBV-type star.  Assumes $\dot M = 2.65
\times 10^{-5} \ M_{\sun} \ {\rm yr^{-1}}$, $a = 2.9$ AU, and
$v_{wind} = 10^3 \ {\rm km \ s^{-1}}$ (see text for details).  Note that
both the maximum amplitude of the absorption and its strong variations
are inconsistent with observations of GRS 1915+105.}
\end{figure}

\begin{figure}
\vspace*{140mm}
\includegraphics{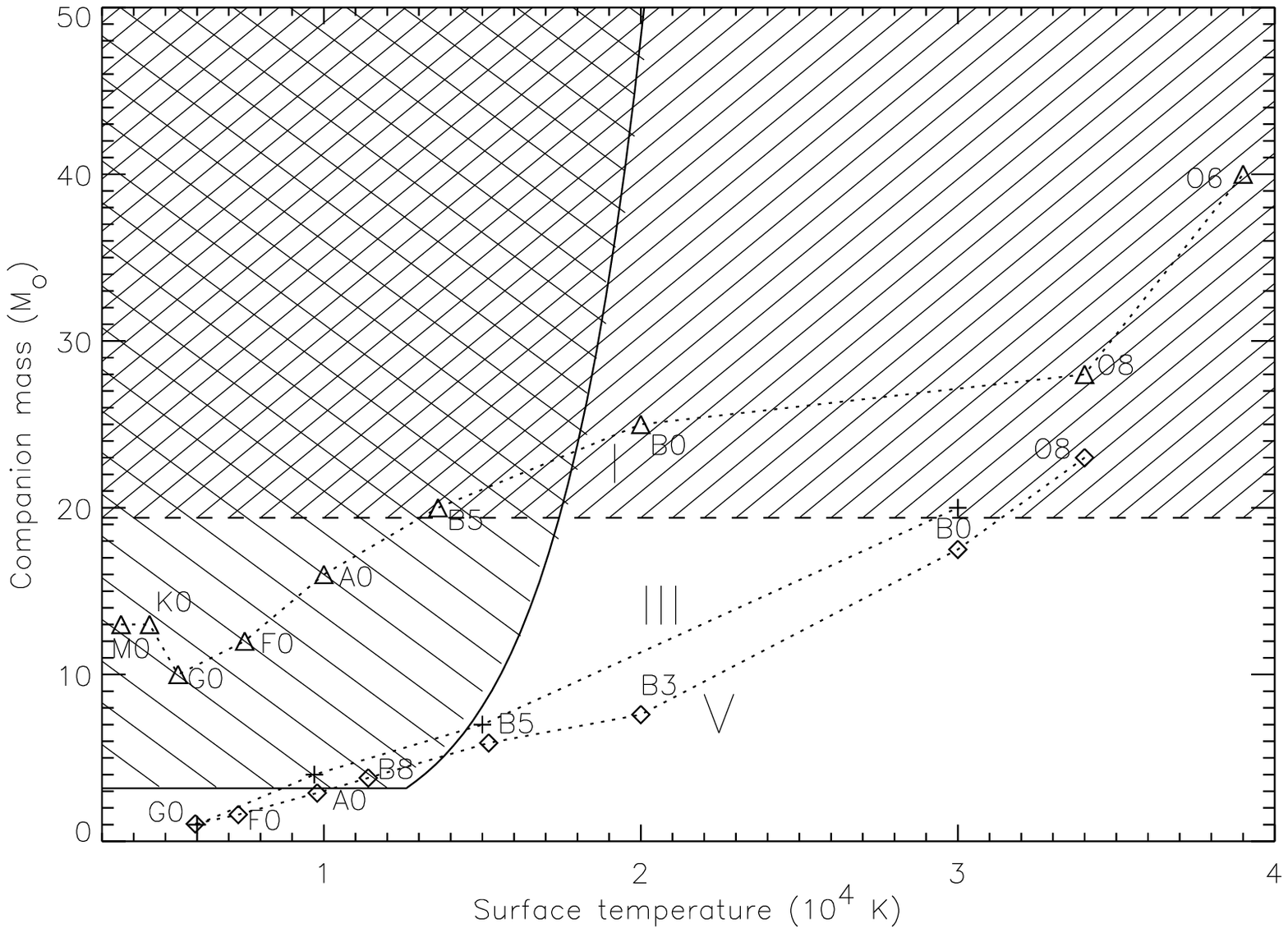}
\caption{Constraints on mass and temperature for stars donating mass
via Roche-lobe overflow.  The hatched areas are regions which are
ruled out by: (i) the limit on $R_s/a$ (bounded by horizontal dashed
line); (ii) limits on X-ray reprocessing (bounded by solid curve);
(iii) the limit on $q > 0.1$ (allowing all stars with $M < 3
M_{\odot}$).  Roman numerals indicate stellar luminosity class for
supergiant (I), giant (III), and main-sequence (V) stars.  Assumes
$M_{BH} = 30 M_{\sun}$.}
\end{figure}

\end{document}